\documentclass[twocolumn,pre,aps,floatfix,superscriptaddress]{revtex4}
\usepackage{amsmath,amssymb,graphicx,float,epstopdf,xcolor}
\begin{document}
\title{Aggregation-fragmentation-diffusion model for trail dynamics}
\author{Kyle Kawagoe}
\affiliation{Kavli Institute for Theoretical Physics, University of California Santa Barbara, CA 93106, USA}
\affiliation{Department of Physics, University of Chicago, Chicago, IL 60637, USA}
\author{Greg Huber}
\affiliation{Kavli Institute for Theoretical Physics, University of California Santa Barbara, CA 93106, USA}
\author{Marc Pradas}
\affiliation{School of Mathematics and Statistics, The Open University, Walton Hall, Milton Keynes, 
MK7 6AA, England}
\affiliation{Kavli Institute for Theoretical Physics, University of California Santa Barbara, CA 93106, USA}
\author{Michael Wilkinson}
\affiliation{School of Mathematics and Statistics, The Open University, Walton Hall, Milton Keynes, 
MK7 6AA, England}
\affiliation{Kavli Institute for Theoretical Physics, University of California Santa Barbara, CA 93106, USA}
\author{Alain Pumir}
\affiliation{Laboratoire de Physique, Ecole Normale Sup\'erieure de Lyon, CNRS, Universit\'e de Lyon, 
F-69007, Lyon, France}
\affiliation{Kavli Institute for Theoretical Physics, University of California Santa Barbara, CA 93106, USA}
\author{Eli Ben-Naim}
\affiliation{Theoretical Division and Center for Nonlinear Studies, Los Alamos National Laboratory, Los Alamos, New Mexico 87545, USA}
\begin{abstract}        
We investigate statistical properties of trails formed by a random
process incorporating aggregation, fragmentation, and diffusion.  In
this stochastic process, which takes place in one spatial dimension,
two neighboring trails may combine to form a larger one and also, one
trail may split into two.  In addition, trails move diffusively.  The
model is defined by two parameters which quantify the fragmentation
rate and the fragment size.  In the long-time limit, the system
reaches a steady state, and our focus is the limiting distribution of
trail weights.  We find that the density of trail weight has power-law
tail $P(w) \sim w^{-\gamma}$ for small weight $w$.  We obtain the
exponent $\gamma$ analytically, and find that it varies continuously
with the two model parameters.  The exponent $\gamma$ can be positive
or negative, so that in one range of parameters small-weight tails are
abundant, and in the complementary range, they are rare.
\end{abstract}
\maketitle

\section{Introduction}
\label{sec-intro}

Processes by which objects may randomly merge or split into smaller
parts are found in a wide range of natural and physical phenomena
\cite{krb,vzl,kr,mkb,bk,lt}, including reversible polymerization
\cite{bt,rm,rdg}, river networks \cite{aes,gh}, and force chains
\cite{lnscmnw,cmnw}. Irreversible aggregation can lead to gelation
where a single aggregate forms and accounts for a finite fraction of
system mass \cite{pjf,aal,zhe,ve,fl}, and irreversible fragmentation
can result in shattering where zero-mass fragments account for a
finite fraction of all mass \cite{afp,mz,cr,es,kb03}. When aggregation
and fragmentation compete, the system typically reaches a steady
state, and the precise balance between merger and breakup controls the
nature of the steady state \cite{mkb,bk,rdg}.

Many studies of aggregation-fragmentation processes do not implicitly
account for aggregate mobility, nor do these models allow for an
underlying spatial structure \cite{krb}.  In this paper we investigate
a stochastic process in which trajectories, which we call ``trails'',
can diffuse, merge, or split.  The trajectories each carry a
``weight'', or population, and the process can serve as a model for
migration trails of animals.  The current investigation complements
recent studies of chaotic dynamics \cite{pphw,hppw} which reveal
distinctive trails which have qualitative and quantitative
similarities to those which are investigated here.

\begin{figure}[t]
\includegraphics[width=0.45\textwidth]{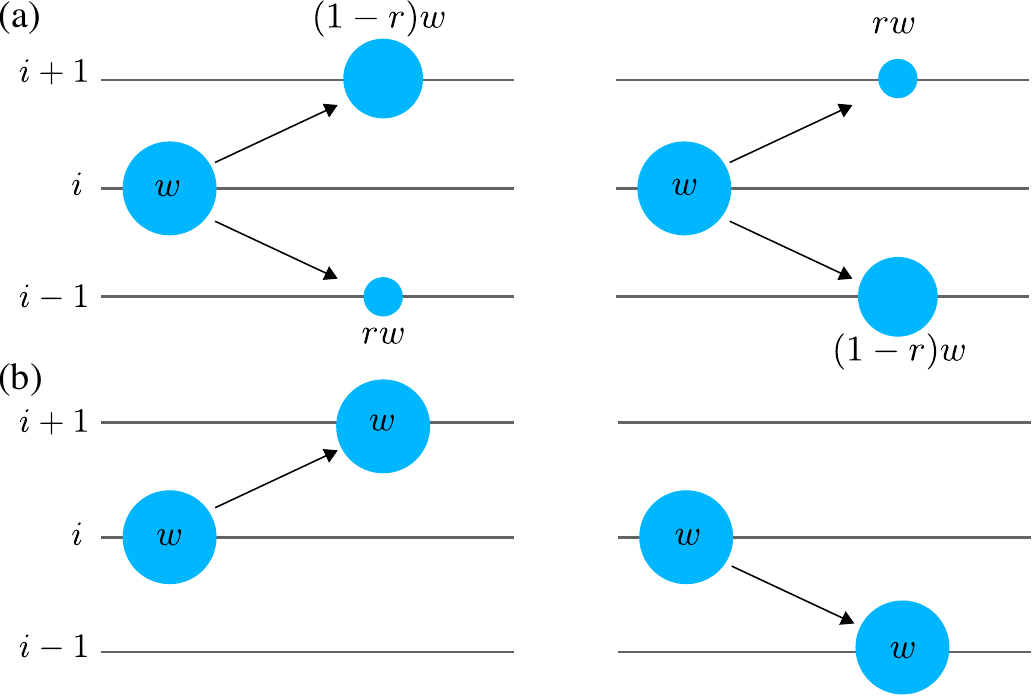}
\caption{Illustration of: (a) fragmentation with probability $p$, with
  a fixed ratio between the two fragments, characterized by $r$, and
  (b) random hopping with probability $1-p$.}
\label{fig-model}
\end{figure}

The stochastic process for the evolution of trails takes place in one
spatial dimension.  In our model, both space and time are discrete. At
each timestep, every trail may split, with probability $p$, into two
smaller trails, or remain intact with the complementary probability
$1-p$ (see figure \ref{fig-model}). In the former case, two trails are
created with weights that are fractions $r$ and $1-r$ of the weight of
the parent trail, and hence, the overall weight is conserved.  Trails
also move randomly, and essentially perform a simple random
walk. Finally, when two trails collide, they merge, with the overall
weight being a conserved quantity.

Regardless of the initial conditions, the system evolves toward a
steady state. We study the steady-state density $P(w)$ of trails with
weight $w$ and find the power-law behavior
\begin{equation}
\label{powerlaw}
P(w) \sim w^{-\gamma},
\end{equation}
for small weights, $w\to 0$. Interestingly, the power-law exponent
varies continuously with the fragmentation probability $p$ and the
fragmentation ratio $r$ as $\gamma$ is a real root of the equation
\begin{equation}
\label{gamma}
r^{\gamma-1}+(1-r)^{\gamma-1}=\frac{3-p}{1-p}\, .
\end{equation}
Results of our numerical simulations are in excellent agreement with
this theoretical prediction. Depending on the parameters $p$ and $r$,
the exponent $\gamma<1$ can be positive such that small-weight trails
are enhanced, or it can be negative such that small-weight trails are
suppressed. Our theoretical approach assumes that spatial correlations
are absent at the steady state, and our extensive numerical
simulations support this assumption.

The rest of this manuscript is organized as follows. Section
\ref{sec-model} introduces the model, and section \ref{sec-density}
provides an elementary derivation of the trail concentration.  In
section \ref{sec-weight-density}, we analyze the density of trail
weights theoretically and numerically.  We obtain analytic expressions
for moments of the trail density, and also, obtain the distribution of
small weights. Section \ref{sec-weak} addresses the weak fragmentation
limit where aggregation and fragmentation proceed independently. In
this limit, dynamical properties of voids between adjacent trails can
be understood using first-passage properties of an ordinary random
walk.  We discuss applications of the model to physical problems, in
particular to trajectories of particles in complex flows in section
\ref{sec-apps}, and conclude in section \ref{sec-conclusions}.

\section{The model}
\label{sec-model}

Our aggregation-fragmentation-diffusion model takes place on an
unbounded lattice in one dimension. A lattice site labeled $i$ may be
either vacant or occupied by a trail with weight $w_i$. The weight can
also be understood as density of a trail of particles concentrated at
location $i$. In the initial configuration, each site is occupied by a
trail with weight unity.

The stochastic process has three elements: (i) {\em
  Fragmentation}. With the probability $p$, a trail with weight $w_i$
splits into two fragments with weights $rw_i$ and $(1-r)w_i$. One of
these fragments moves to neighboring site $i-1$ and the remaining
fragment moves to neighboring site $i+1$.  The two realizations are
equally likely (see figure \ref{fig-model}).  (ii) {\em Diffusion}:
With the complementary probability $1-p$ the trail remains intact and
it moves, with equal probabilities, to site $i-1$ or to site $i+1$.
(iii) {\em Aggregation}. All sites are updated simultaneously
according to the fragmentation and diffusion steps above.  When two
distinct trails arrive at the same site, they immediately merge to
form a new trail whose weight equals the sum of those of the two
original trails.

\begin{figure}[t]
\includegraphics[width=0.47\textwidth]{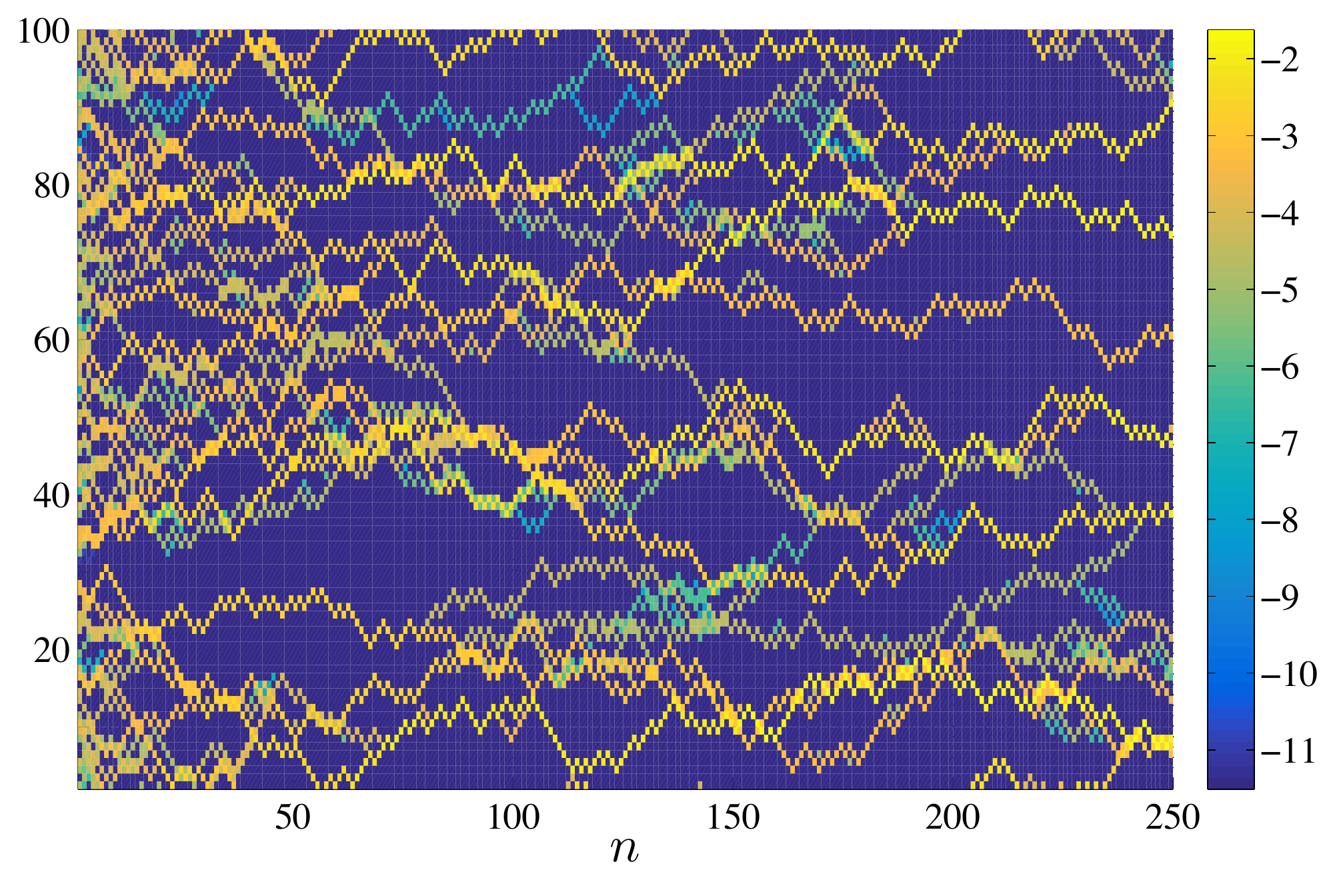}
\caption{A numerical realization of the
  aggregation-fragmentation-diffusion process with $r=0.01$ and
  $p=0.05$. The color coding illustrates weights from low (blue) to
  high (yellow) values, and the color bar is in natural logarithmic
  scale.  Unoccupied sites are dark purple.}
\label{fig-spacetime}
\end{figure}

Hence, two parameters characterize the model: the fragmentation
probability, $p$ with $0<p<1$, and the fragmentation ratio, $r$ with
$0<r<1$.  These parameters control how the weight at each site evolves
prior to the final aggregation step,
\begin{equation}
w\to 
\begin{cases}
\label{process}
rw,(1-r)w & \quad{\rm with\ prob.\ } p\, , \\
w         & \quad{\rm with\ prob.\ } 1-p\, .
\end{cases}
\end{equation}
All trails are updated simultaneously, and time $n$ is augmented by
one after each iteration, $n\to n+1$.  Essentially, trails perform a
random walk (see figure \ref{fig-spacetime}), and importantly,
mobility is not coupled to weight.  We stress that total trail weight
is conserved since both fragmentation and aggregation are conservative
processes.

Our Monte Carlo simulations were performed using a regular lattice
with $N$ sites and periodic boundary conditions. Initially, each site
is occupied with a trail of unit weight.  In each iteration, all sites
are updated simultaneously according to the model rules.  Namely, a
trail jumps without splitting to a neighboring site with probability
$1-p$, or decomposes into two fragments, with probability $p$, as in
figure \ref{fig-model} and equation \eqref{process}. Two trails
landing at iteration $n$ at the same location immediately
merge. Figure \ref{fig-spacetime} shows trajectories in a space-time
diagram using simulation data.
 
\section{The concentration}
\label{sec-density}

We first study the trail concentration $c$, defined as fraction of
occupied sites.  Our primary focus is the steady state, where the
competing processes of aggregation and fragmentation balance each
other.  The fragmentation ratio $r$ affects the trail weight but it
does not affect the number of trails.  Hence, the concentration
depends on the fragmentation probability $p$ alone. By assuming that
occupations at neighboring sites are not correlated, we can write a
closed equation for the concentration.  In each fragmentation event a
single trail generates two trails and conversely, in each aggregation
event two trails coalesce into one.  At the steady state, the gain
rate and the loss rate balance,
\begin{equation}
\label{balance}
p\,c=\left(\frac{1+p}{2}c\right)^2\,.
\end{equation}
The fragmentation rate on the left-hand-side is proportional to the
concentration $c$ and the fragmentation probability $p$. The
aggregation rate equals the probability that two trails arrive at the
same site.  The quantity in parentheses is the sum of $p\,c$ and
$\frac{1-p}{2}c$ accounting for trail fragments and intact trails
respectively (see figure \ref{fig-model}).  In writing the quadratic
aggregation term in \eqref{balance}, we make the assumption that the
occupancy at a site is not correlated with that at its next-nearest
neighbor.

\begin{figure}[t]
\includegraphics[width=0.47\textwidth]{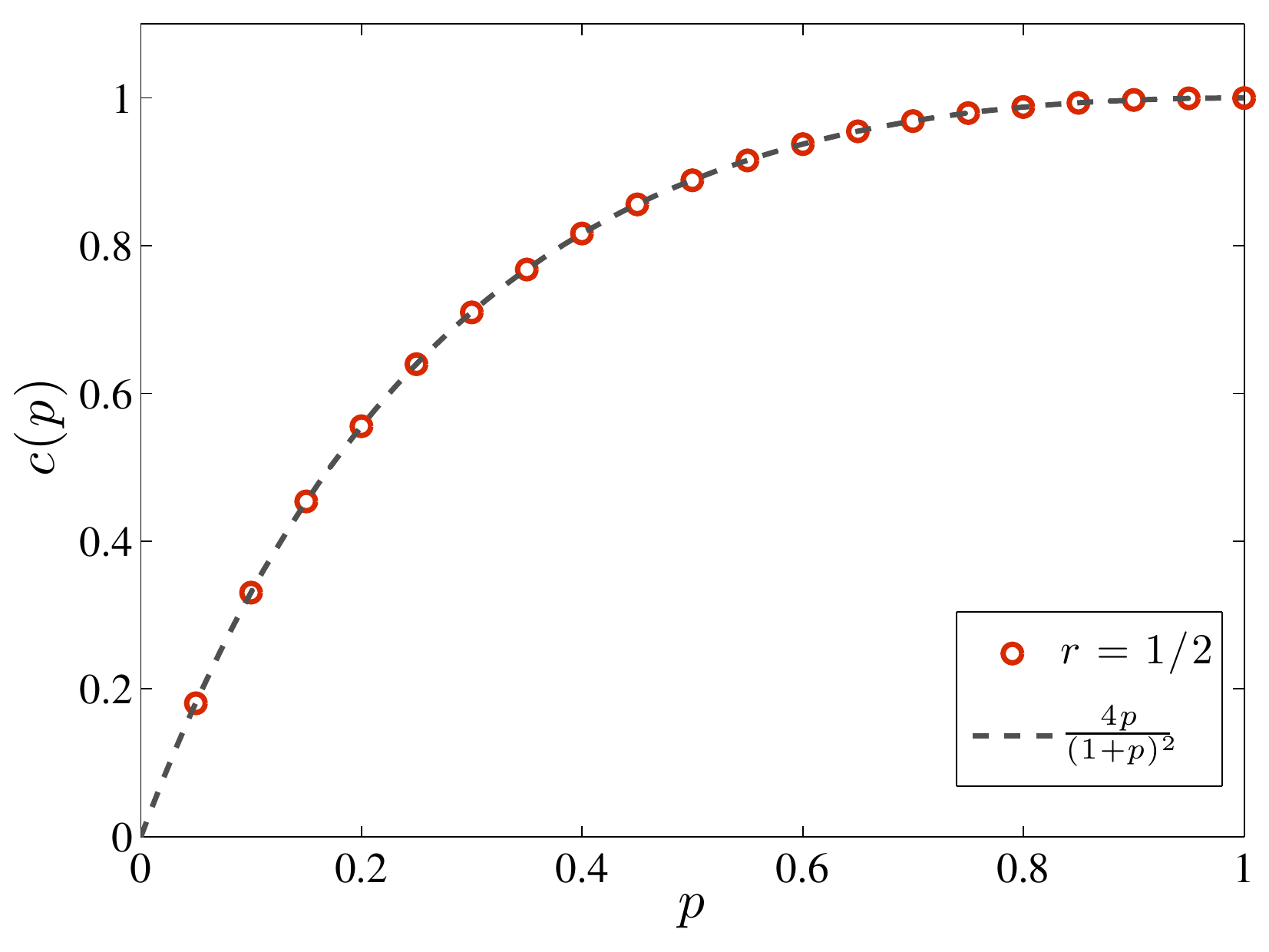}
\caption{Numerically computed trail concentration $c(p)$ versus the
  fragmentation probability $p$.  The dashed line corresponds to the
  analytic expression \eqref{c}, and circles to Monte Carlo
  simulations on a lattice of $N=10^6$ sites.}
\label{fig-cp}
\end{figure}

Rearranging Eq.~\eqref{balance} we find the trail concentration 
\begin{equation}
\label{c}
c=\frac{4p}{(1+p)^2}\,. 
\end{equation}
Figure \ref{fig-cp} shows a comparison of Eq.~\eqref{c} with numerical
simulations of the model.  The numerical results indicate that spatial
correlations in the trail concentration disappear in the steady state.
Qualitatively, the equilibrium concentration of domain walls in the
one-dimensional Ising model with single-spin flip dynamics exhibits
similar phenomenology \cite{rjg,zr,gns}.
 
The trail concentration has the following limiting behaviors 
\begin{equation}
\label{c-limits}
c\simeq 
\begin{cases}
4p & \quad p\to 0 \,,\\
1-\frac{1}{4}(1-p)^2 & \quad p\to 1\,.
\end{cases}
\end{equation}
The concentration vanishes linearly when $p\to 0$, and hence, the
average trail weight which according to mass conservation is inversely
proportional to $c$, diverges in this limit. Also, the fraction of
vacant sites vanishes quadratically when $p\to 1$.

\section{The weight density}
\label{sec-weight-density}

We now turn to the main focus of our investigation, the steady-state
weight density.  Our theory builds on the results of section
\ref{sec-density}, and using the assumption that weights at different
sites are not correlated, we can accurately predict key statistical
properties of the weight density.

We define the weight density $P(w)$ such that $P(w)dw$ is the fraction
of sites occupied by a trail with weight in the infinitesimal range
$[w:w+dw]$.  Trail weight is conserved throughout the
aggregation-fragmentation-diffusion process and hence, the total
weight density $\int dw\,w\, P(w)$ is a constant.  The concentration
$c$ equals the integrated weight density
\begin{equation}
\label{c-int}
c=\int_0^\infty dw\, P(w)\,.
\end{equation}
The normalized quantity $c^{-1}P(w)$ is the probability distribution
function for the weight.

Changes in trail weight occur in two stages: first, before trails move
and then, after they move.  In the first stage, trail weight may
change by fragmentation. Let $G(w)$ be the weight density of trails
produced at the first stage.  According to the random process
\eqref{process}, we have
\begin{equation}
\label{Gw-def}
G(w)=\frac{p}{r}P\left(\frac{w}{r}\right)+\frac{p}{1-r}P\left(\frac{w}{1-r}\right)+(1-p)P(w)\, .
\end{equation}
Here, the first two terms on the right-hand side account for fragments
and the last term accounts for intact trails. As also follows from
Eq.~\eqref{Gw-def}, weight conservation sets \hbox{$\int dw\,w\,
  G(w)=\int dw\,w\, P(w)$}, and the fragmentation rule \eqref{process}
implies \hbox{$\int dw\, G(w)=(1+p)\int dw\,P(w)$}.

In the steady state, the trail density before an iteration, $P(w)$, is
unchanged after one iteration of the combined
aggregation-fragmentation diffusion process. This is expressed by the
nonlinear-integral equation:
\begin{eqnarray}
\label{Pw-eq}
P(w)=\left(\!1-\frac{1+p}{2}c\!\right)\!G(w)+\frac{1}{4}\int_0^w \!\!dv\, 
G(v)G(w-v) .
\end{eqnarray}
The first term on the right-hand side accounts for the scenario where
there is no aggregation. It is a product of three factors: (i) The
quantity $G(w)/2$ which represents a trail produced at a neighboring
site and where the factor $1/2$ accounts for the equal distribution of
weight from one site to its two neighbors, (ii) The factor $2$
accounting for two neighbors, and (iii) The probability
$1-\tfrac{1+p}{2}c$ that such a trail avoids aggregation.  This
probability sums $1-c$ and $\tfrac{1-p}{2}c$ for a vacant and an
occupied next-nearest neighbor, respectively.  The second term on the
right-hand side is the aggregation term; it is a convolution of two
identical terms of the form $G(w)/2$ with $G(w)$ given by
\eqref{Gw-def}.  By integrating Eq.~\eqref{Pw-eq} over all weights, we
recover equation \eqref{balance}, and furthermore, this equation is
consistent with mass conservation.

We stress that substitution of equation \eqref{Gw-def} into
\eqref{Pw-eq} turns the latter into a closed, nonlinear, equation for
the weight density $P(w)$. Compared with Eq.~\eqref{balance}, the
steady-state equation \eqref{Gw-def} makes an even stronger assumption
that {\em weights} at different sites are not correlated. Indeed, the
convolution term in \eqref{Pw-eq} which accounts for the weight of
aggregates is quadratic in the density $P(w)$.

\subsection{Moment Analysis}

To examine the validity of this no-correlation assumption, we
study the moments of the weight density,
\begin{equation}
\label{Mk-def}
M_k=\int dw\, w^k P(w)\,.
\end{equation}
Of course, the zeroth moment corresponds to the total concentration
$M_0=c$, and the first moment $M_1=1$ corresponds to the total weight
density which is a conserved quantity. As shown in Fig.~\ref{fig-cp},
the numerical simulations confirm the predictions of \eqref{Pw-eq} for
the zeroth moment.

\begin{figure}[t]
\includegraphics[width=0.47\textwidth]{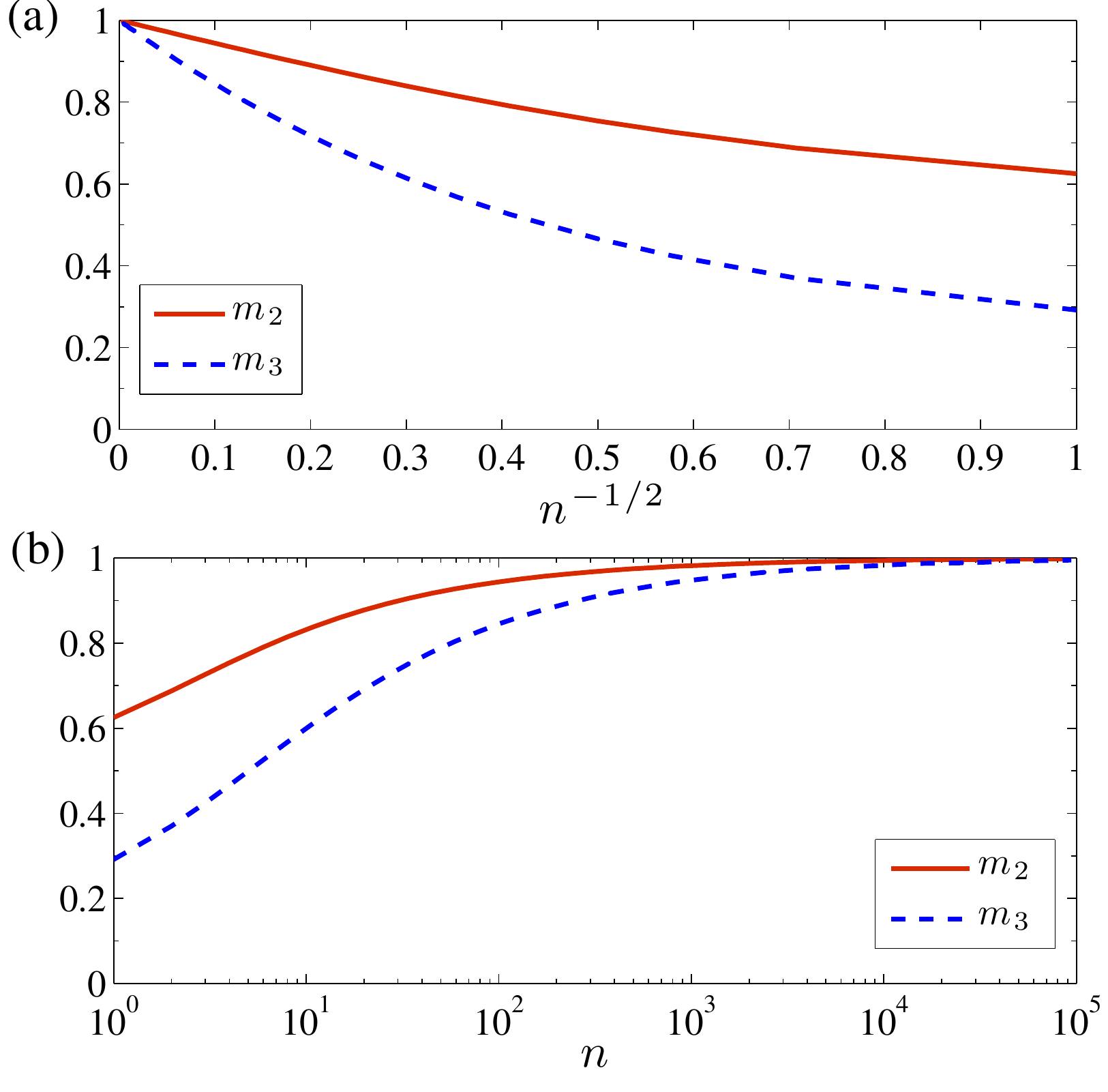}
\caption{The normalized second and third moments $m_2(n)=M_2(n)/2$
  and $m_3(n)=M_3(n)/6$, versus: (a) $n$, and (b) $n^{-1/2}$.  Shown
  are results of Monte Carlo simulations on a ring with $10^8$ sites
  for the  case $r=p=1/2$. A fourth order polynomial fit to
  $m_k(n)$ versus $n^{-1/2}$ yields the asymptotic values $1.0001$ and
  $1.0002$ for $m_2(\infty)$ and $m_3(\infty)$.}
\label{fig-m23}
\end{figure}

We now introduce the Laplace transform \hbox{$M(s)=\int_0^\infty dw
  e^{-sw} P(w)$} which is the generating function of the moments,
\begin{equation}
\label{Ms-def}
M(s)=\sum_{n=0}^\infty \frac{(-s)^k}{k!}M_k\,.
\end{equation}
By first substituting the concentration \eqref{c} into \eqref{Pw-eq}, then
multiplying equation \eqref{Pw-eq} with $e^{-sw}$, and finally,
integrating over weight, we find that the Laplace transform obeys
the nonlinear equation
\begin{equation}
\label{Ms-eq}
M(s)=\frac{1-p}{1+p}\,U(s)+\frac{1}{4}U^2(s)\, ,
\end{equation}
with \hbox{$U(s)=pM(rs)+pM(s-rs)+(1-p)M(s)$}. The quantity $U(s)$ is
the following Laplace transform \hbox{$U(s)=\int dw e^{-sw} G(w)$},
and it can be conveniently expressed in terms of the moments $M_k$,
\begin{equation}
\label{Us-def}
U(s)=\sum_{n=0}^\infty  \frac{(-s)^k}{k!} \,u_k\,M_k\,.
\end{equation}
Here, \hbox{$u_k=1-p[1-r^k-(1-r)^k]$}, and we quote the values
$u_0=1+p$, $u_1=1$, and $u_2=1-2pr(1-r)$.  Using $M(0)=c$ and
$U(0)=(1+p)c$, we can recover from \eqref{Ms-eq} the trail
concentration \eqref{c}.  Mass conservation dictates that $M_1$ equals
the initial mass density. In general, the moments satisfy the
recursion
\begin{equation}
\label{Mk-rec}
M_k=\frac{1}{4(1-u_k)}\sum_{l=1}^{k-1}\binom{k}{l}u_lu_{k-l}M_lM_{k-l}\,,
\end{equation}
when $k\geq 2$. In particular, the second and third moments are given
by
\begin{equation}
\label{M23}
M_2=\frac{M_1^2}{2(1-u_2)}\,,\quad M_3=\frac{3\,u_2\,M_1^3}{4(1-u_2)(1-u_3)}\,.
\end{equation}
For the special case $p=r=1/2$ with $M_1=1$, we have $M_0=8/9$,
$M_2=2$ and $M_3=6$. Results of our Monte Carlo simulations are in
excellent agreement with these theoretical predictions (see figure
\ref{fig-m23}). We also verified numerically that \eqref{M23} holds
for other values of the fragmentation probability $p$ and the
fragmentation ratio $r$, and for a variety of initial conditions.

\begin{figure}[t]
\includegraphics[width=0.47\textwidth]{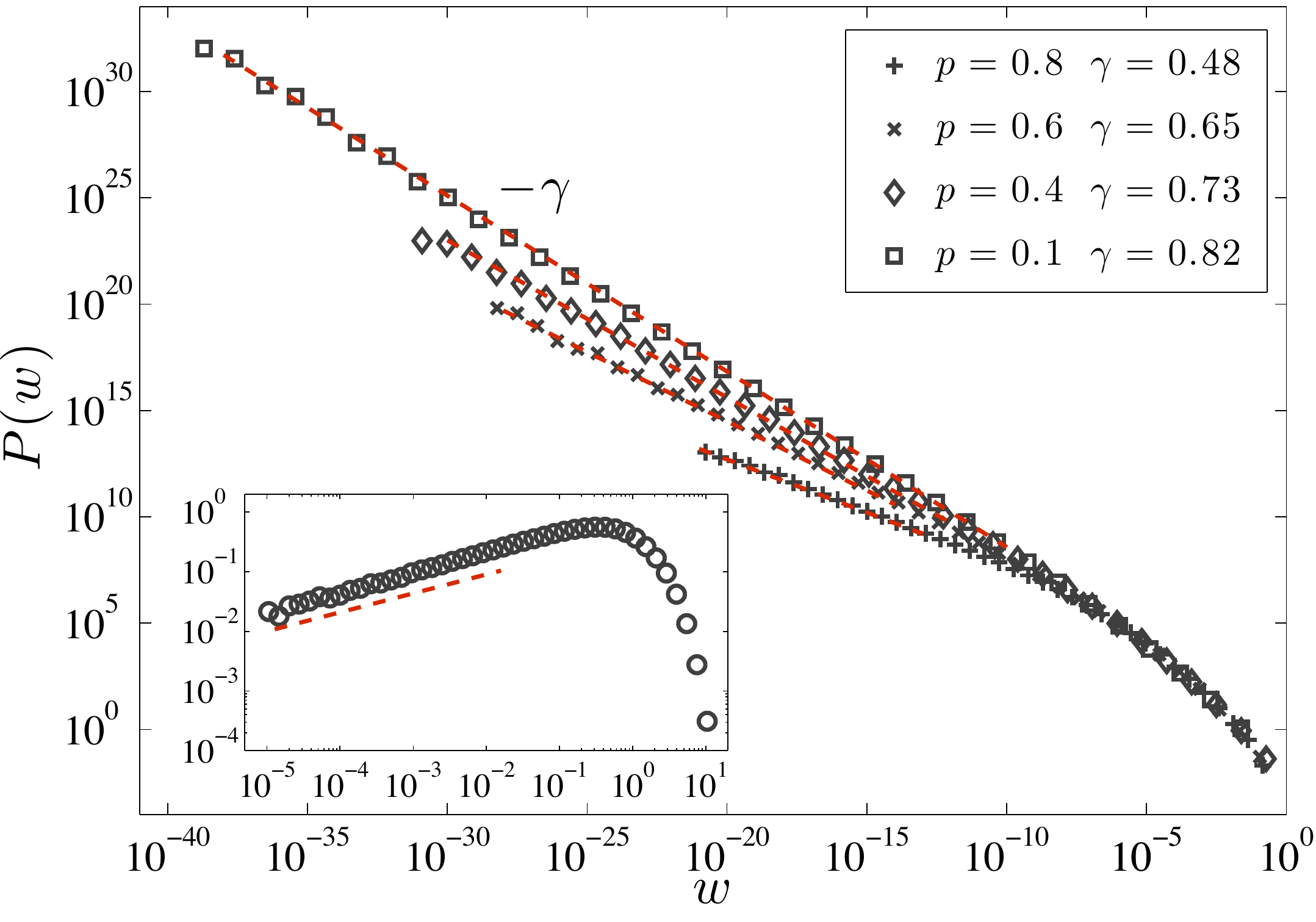}
\caption{The power-law tail of the weight density $P(w)$.  The curves
  shown here correspond to different values of $p$, indicated in the
  legend, and are obtained from Monte Carlo simulations on a lattice
  of $N=10^6$ sites. Dashed lines correspond to the theoretical
  predictions of equations \eqref{powerlaw}-\eqref{gamma}.  The value
  of $r$ was chosen to be $r = 0.01$. The inset
    displays the case $r=0.25$ and $p=0.7$ for which $\gamma=-0.32$.}
\label{fig-Pw}
\end{figure}

We have also examined numerically the empty-interval probability $E_l$
that $l$ consecutive sites are vacant. As expected given the absence
of spatial correlations, we confirmed the exponential behavior
$E_l=E_1^l$ with $E_1=1-c$ and $c$ given in \eqref{c}.  Our
preliminary analytic results for the simpler case of sequential
dynamics where sites are updated one at a time show that spatial
correlations do not vanish, and therefore, qualitative features of the
steady state are sensitive to details of the dynamics.

Below, we discuss the small-weight statistics of the trail density and
provide numerical results that confirm the theoretical
predictions. Based on the simulation results, we conclude that at the
steady state, spatial correlations in trail weight disappear.  A
similar behavior occurs in adsorption-desorption processes
\cite{pb,kb,jtt} where gaps between adsorbed particles in
one-dimension essentially undergo an aggregation-fragmentation
process, and also, in a related model for the propagation of force
chains in granular matter \cite{cmnw}.

\begin{figure}[t]
\includegraphics[width=0.47\textwidth]{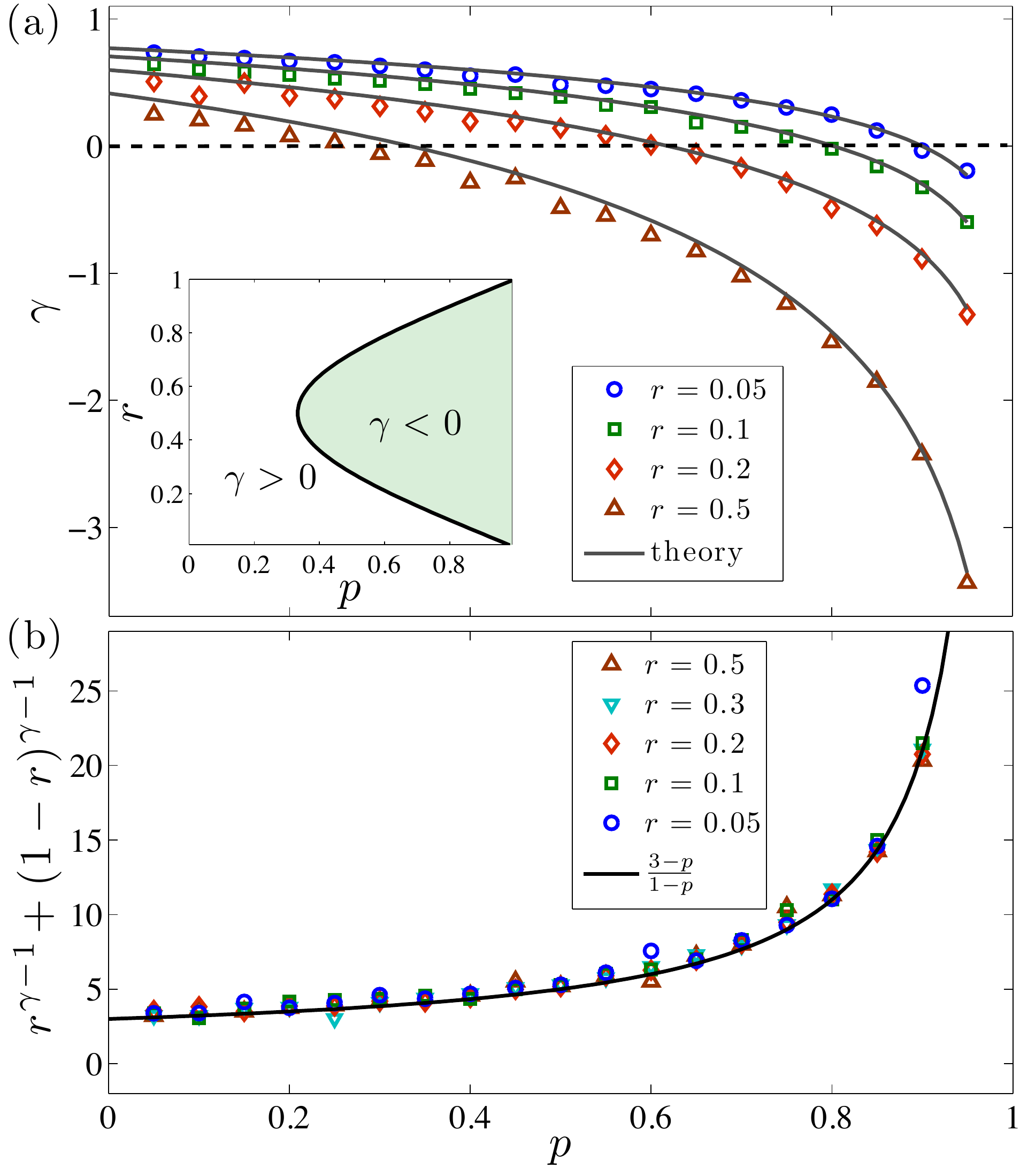}
\caption{(a) The exponent $\gamma$ for different values of $p$ and
  $r$. The lines correspond to the theoretical predictions, given by
  Eq.~\eqref{gamma}, for different values of $r$, see the legend. The
  inset shows the region of negative $\gamma$ as shaded.  (b) When
  data is plotted as $r^{\gamma-1} + (1-r)^{\gamma-1}$ it collapses
  into the curve $(3-p)/(1-p)$ (solid line) in accord with
  \eqref{gamma}.}
\label{fig-gp}
\end{figure}

The results of section \ref{sec-density} imply that in a finite
system, every configuration with $N_1$ occupied sites and $N_2$ vacant
sites is realized with the equilibrium probability
$c^{N_1}(1-c)^{N_2}$ that corresponds to a noninteracting two-state
system. The moment analysis indicates that an even stronger statement
applies as all states with the same configuration of weights are
equally probable. In essence, the weight density $P(w)$ completely
describes the system.

\subsection{Small-weight statistics}

We now focus on the small-weight tail of the density $P(w)$.  In the
limit $w\to 0$, the nonlinear terms become negligible.  If we
substitute the concentration given in \eqref{c} into \eqref{Pw-eq}, we
find that the small-$w$ tail of the steady-state density $P(w)$
satisfies the {\em linear} equation
\begin{eqnarray}
\label{Pw-tail-eq}
\frac{1}{r}P\left(\frac{w}{r}\right)+\frac{1}{1-r}P\left(\frac{w}{1-r}\right)=
\frac{3-p}{1-p}P(w)\,.
\end{eqnarray}
From this equation we arrive at our main result, the power-law
behavior $P(w)\sim w^{-\gamma}$ with the exponent $\gamma$ being root
of Eq.~\eqref{gamma}. Clearly, the exponent $\gamma<1$ varies
continuously with the model parameters $r$ and $p$. In the special
case $r=1/2$ we have $\gamma=2-(\ln \tfrac{3-p}{1-p})/\ln 2$.

\begin{figure}[t]
\includegraphics[width=0.47\textwidth]{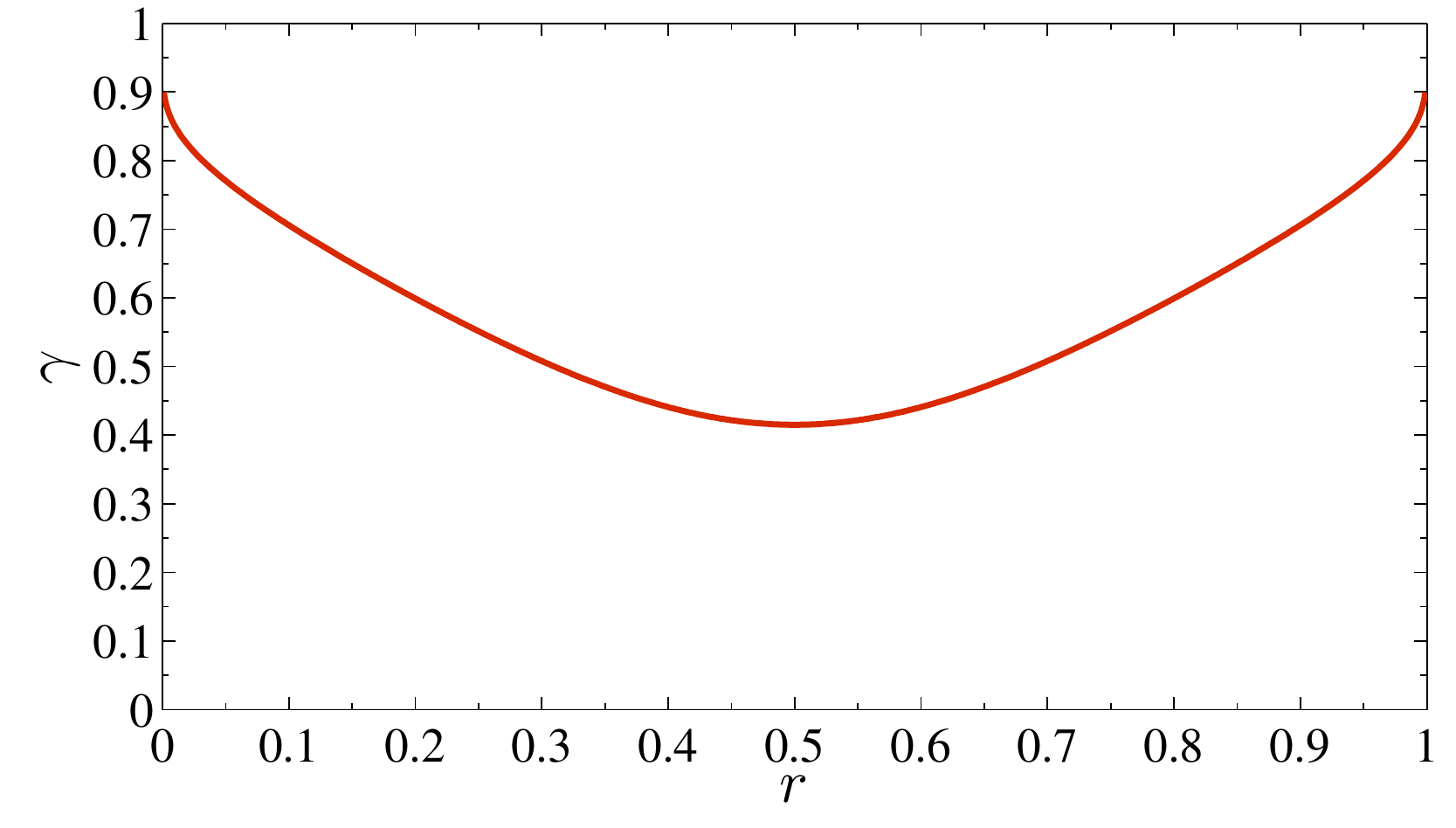}
\caption{The exponent $\gamma$ versus the parameter $r$ for small 
  values of $p$.}
\label{fig-gamma}
\end{figure}

The small-weight statistics also follow from the Laplace
transform. When $M(s)$ is small, the nonlinear term in the governing
equation \eqref{Ms-eq} is negligible, thereby leading to the linear equation
$(1+p)M(s)=(1-p)U(s)$.  This equation implies the large-$s$ decay
$M(s)\sim s^{\gamma-1}$, which is equivalent to the power-law behavior
\eqref{powerlaw}, with the exponent $\gamma$ root of equation
\eqref{gamma}.

\label{sec-weak}
   \begin{figure*}[t]
\includegraphics[width=0.96\textwidth]{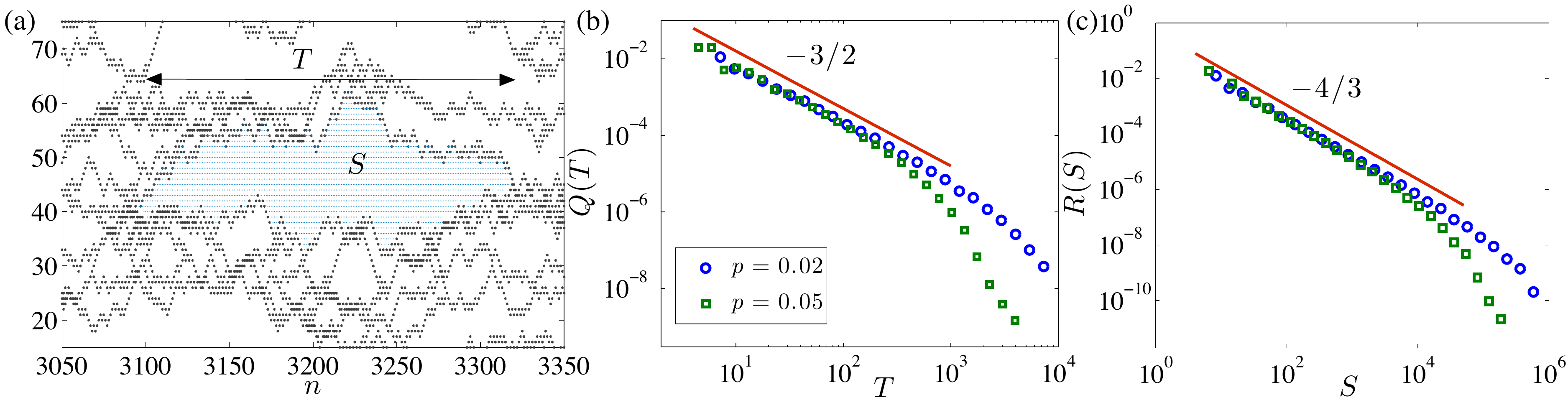}
\caption{(a) Voids are defined as closed regions of unoccupied sites
  and they are defined by an area $S$ (blue region) and duration
  $T$. (b) and (c) PDF of the void durations and sizes, respectively,
  for two different values of $p$ and $r=0.01$.  Solid red lines correspond to
  a power law with exponent $-3/2$ (b) and $-4/3$ (c).}
\label{fig-lac}
\end{figure*}

Our numerical simulations provide excellent support for equation
\eqref{gamma}.  Figure \ref{fig-Pw} demonstrates the power-law tail of
the weight density, and figure \ref{fig-gp} compares the
numerically-computed exponent $\gamma$ for different model parameters
with the theoretical prediction.

In a finite system of size $N$, the power-law tail \eqref{powerlaw}
holds in the range $N^{-1/(1-\gamma)}\ll w \ll 1$. The lower limit
follows from the criterion $N\int_0^z dw\,P(w)\sim 1$ which gives an
estimate for the scale $z$ of the smallest weight in the system
\cite{krb}.  The scaling law $z\sim N^{-1/(1-\gamma)}$ explains the
large variation in the extent of the power-law regime observed in
figure \ref{fig-Pw}.

Equation \eqref{gamma} implies that there are two distinct regimes of
behavior, since the exponent $\gamma$ vanishes when
\begin{equation}
\label{vanish}
r(1-r)=\frac{1-p}{3-p}\,.
\end{equation}
As shown in figure \ref{fig-gp}a, $\gamma$ can be positive, in which
case, the density of small-weight tails is enhanced. In the
complementary regime, $\gamma$ is negative and small-weight trails are
suppressed. When $p<1/3$ the exponent is always positive, $\gamma>0$.

The linear equation \eqref{Pw-tail-eq} reflects the nature of the
fragmentation process in which weight ``cascades'' from small trails
into even smaller trails according to \hbox{$w\to rw,(1-r)w$}\,.  This
cascade process is balanced by aggregation of small trails into larger
trails. The power-law behavior is valid over a substantial size range,
and throughout this range only terms that are linear in $P(w)$
dominate Eq.~\eqref{Pw-eq}. Qualitatively similar cascades, where the
full nonlinear theory reduces to a linear theory for extreme-value
statistics, are found in wave turbulence \cite{crz,zlf} and inelastic
gases \cite{bm}. Yet, the cascade process described by equation
\eqref{Pw-tail-eq} has two distinctive features. First, the tail
\eqref{powerlaw} can be vanishing or diverging. Second, in equation
\eqref{Pw-eq}, the term linear in $P(w)$ is also proportional to the
trail concentration $c$. Consequently, the prefactor on the right-hand
side of \eqref{Pw-tail-eq} depends on the steady-state value \eqref{c}
of the concentration, and despite its linear nature, this equation
does incorporate a two-point correlation.

For completeness, we mention that we also numerically studied the
large-$w$ tail of the weight density. In contrast with the broad
power-law tail that may occur at small weights, we find that large
weights are exponentially rare, $P(w)\sim \exp(-{\rm const.}\times w)$
for $w\gg 1$. This behavior is consistent with the large-size
statistics found in irreversible aggregation \cite{fl} and in the
closely related $q$-model for force chains \cite{lnscmnw,cmnw}.

\subsection{Stochastic Fragmentation}

The fragmentation process in \eqref{process} is deterministic in the
sense that the sizes of the two trail fragments are fixed fractions of
the original trail. We briefly mention a natural counterpart,
stochastic fragmentation, where the fraction $0<r<1$ is drawn from the
distribution $\eta(r)$. We require that the distribution be: (i)
normalized $\int_0^1 dr \eta(r)=1$, and (ii) symmetric
$\eta(r)=\eta(1-r)$. It is straightforward to generalize the above
theoretical analysis to stochastic fragmentation and in particular,
equation \eqref{gamma} becomes
\begin{equation}
\label{gamma1}
\int_0^1 d\eta\, \eta(r) \left[r^{\gamma-1}+(1-r)^{\gamma-1}\right]=\frac{3-p}{1-p}\, .
\end{equation}
For the so-called random-scission model \cite{krb,cmnw} where the
fraction $r$ is uniformly distributed, $\eta(r)=1$, we find
\begin{equation}
\label{gamma2}
\gamma=2\,\frac{1-p}{3-p}\,.
\end{equation}
In this case, the exponent $\gamma$ satisfies $0<\gamma<2/3$, and the
small-weight tail of the distribution is enhanced, regardless of $p$.

\section{Weak Fragmentation}
\label{sec-weak}

We now analyze the small $p$ case and show that aggregation and
fragmentation proceed independently in this regime.  In the limit
$p\to 0$, the trail concentration vanishes, see Eq.~\eqref{c}, and
consequently, $P(w)\propto p$. By keeping dominant terms of the order
${\cal O}(p^2)$ and neglecting sub-dominant terms of the orders ${\cal
  O}(p^3)$ and ${\cal O}(p^4)$, we find that equation \eqref{Pw-eq}
simplifies to
\begin{eqnarray}
\label{Pw-eq-1}
0
&=&
4p\left[
\frac{1}{r}P\left(\frac{w}{r}\right)+
\frac{1}{1-r}P\left(\frac{w}{1-r}\right)
-P(w)\right]\nonumber\\
&+&
\int_0^w {\rm d}y\ P(y)P(w-y)-2\,c\,P(w)\, .
\end{eqnarray}
The first two (linear) terms account for fragmentation, while the next
two (nonlinear) terms account for aggregation between two intact
trails.  Thus, fragmentation and aggregation are not coupled when
fragmentation is weak.  As a check of self-consistency, we integrate
\eqref{Pw-eq-1} to find $c^2=4p\,c$. Hence, the concentration is
proportional to the fragmentation probability $c\simeq 4p$ as in
\eqref{c-limits}.

In the limit $w\to 0$, the nonlinear term in Eq.~(\ref{Pw-eq-1}) is
negligible, and the linear terms satisfy 
\begin{eqnarray}
\label{P-eq-smallx}
\frac{1}{r}P\left(\frac{w}{r}\right)+
\frac{1}{1-r}P\left(\frac{w}{1-r}\right)=3P(w)\,.
\end{eqnarray}
This linear equation accepts a solution of algebraic form $P(w) \sim
w^{-\gamma}$ for small $w$, where the exponent $\gamma<1$ is root
of the equation
\begin{equation}
\label{gamma-smallp}
r^{\gamma-1}+(1-r)^{\gamma-1}=3.
\end{equation}
Figure \ref{fig-gamma} shows a plot of $\gamma(r)$ as given by the 
above equation where we observe that the minimal value
$\gamma=2-\frac{\ln 3}{\ln 2}$ is attained when $r=1/2$.

As discussed above, when fragmentation is weak, the trail
concentration is low. Moreover, since fragmentation events are rare,
fragmentation and aggregation proceed independently. In this limit, it
is relatively simple to characterize the dynamics of voids between
adjacent trails. In the limit $p\to 0$, trails merely perform a random
walk, and when two trails come to within distance of two sites, there
is a finite probability for the two to coalesce. Hence, the problem is
equivalent to a first-passage process of a simple random walk (the
distance between two independent random walks itself performs a random
walk). Let $T$ be the lifetime of a void between two trails and $Q(T)$
be the probability a void has a lifetime $T$ (see figure
\ref{fig-lac}a). Using the well-known return probability of a random
walk \cite{sr}, we conclude (see figure \ref{fig-lac}b)
\begin{equation}
\label{QT}
Q(T)\sim T^{-3/2}\,.
\end{equation}
Furthermore, we can also consider the area $S$ of a void in a
space-time diagram (see figure \ref{fig-lac}a). Because the edges of
the void are random walks, the width of the void scales as $T^{1/2}$
and consequently the area scales as $S\sim T^{3/2}$. Using $R(S)$ for
the probability of observing a void with area $S$, the 
scaling behavior (see figure \ref{fig-lac}c)  
\begin{equation}
\label{RS}
R(S)\sim S^{-4/3}\,,
\end{equation}
follows immediately from \eqref{QT}. Our numerical simulation results,
shown in figures \ref{fig-lac}b and \ref{fig-lac}c, support the
first-passage behaviors \eqref{QT} and \eqref{RS}.

\section{Applications}
\label{sec-apps}

The aggregation-fragmentation-diffusion process can be used to model
the migration trails of animals \cite{bm1,ra}. In this application the
time variable is viewed as a second space coordinate, along which the
animal migration is directed, for example north to south. Movement
along the transverse direction is random.  The migrating animals are
assumed to choose the easiest path with a component in their intended
direction of travel.  In a random landscape, the trails will fluctuate
in direction, and different trails may combine. Fragmentation would
occur if there is some inhomogeneity in the population, so that there
may be a difference in which is the easiest path for different
individuals. For example, in human migration, people traveling on foot
may be able to follow a trail which would not be accessible in a
wheeled vehicle.

Another motivation for our study comes from a theoretical model of
particles in a compressible turbulent flow in one dimension of space
\cite{rg,mr,fgv,tb,gm,pw}.  Specifically, consider a large number of
trajectories, in a system with periodic boundary conditions, evolving
according the dynamical system defined in \cite{pphw,hppw}. Assuming
that the trajectories are initially uniformly distributed in space,
trajectories tend to aggregate and form strongly concentrated
clusters, leading to an extremely heterogeneous distribution of
``mass'' in space and time. The merging and splitting particle
trajectories \cite{pphw,hppw} share many qualitative similarities with
those shown in Figure \ref{fig-spacetime}.

The analogy is in fact more than qualitative. The existence of large
regions, with an extremely small concentration of particles is
characterized in the particle problem, and in the
aggregation-fragmentation-diffusion model in the weak $p$ limit, by a
power-law behavior of the area $S$ of the empty regions in a
space-time diagram, defined in Fig.~\ref{fig-lac}a, with an exponent
$4/3$ \cite{pphw}. This points to similar mechanisms to explain the
large heterogeneities in both models.  Furthermore, dividing space
into small bins, and defining the weight $w$ as the number of
trajectories per bin at a given instant of time, one finds a power-law
distribution of $w$ as in Eq.~\eqref{powerlaw} \cite{pphw}. Over the
range of control parameters in the model, the exponent $\gamma$ is
always positive, implying a very large probability of empty regions.
It would be interesting to explore further analogies between the
present model of aggregation-fragmentation-diffusion, and the
theoretical model for particle transport by compressible turbulent
flows. The resulting theoretical understanding should provide insight
on experimental observations on the dispersion of particles on a
surface flow \cite{lbgp}.

\section{Discussion}
\label{sec-conclusions}

We have studied an aggregation-fragmentation-diffusion random process
that describes the evolution of trails of particles with local weight
density. In our model, a trail may fragment into two or it may stay
intact. In addition, trails move randomly, essentially performing
diffusion. Aggregation occurs when two trails arrive at the same
location. The model is characterized by two parameters, the
fragmentation probability which controls the relative strength of the
fragmentation process, and the fragmentation ratio which controls the
size of the produced fragments.  An equilibrium state is found, where
the two competing processes of aggregation and fragmentation balance
each other. In this steady state, the small-weight tail of the
fragment density has a power-law tail. The exponent governing this
tail varies continuously with the model parameters.

At the core of our theoretical approach are the assumptions that the
occupancy and even the weights of trails at different locations are
uncorrelated. Our extensive numerical simulations confirm this
behavior.  The system achieves an equilibrium state, where aggregation
and fragmentation are in perfect balance, with the remarkable property
that all configurations with the same number of trails are equally
probable. Usually, it is possible to trace such behavior to a detailed
balance condition where there is zero net flux between any two
microscopic configurations of the system. It is an interesting
challenge to construct an equivalent condition for our synchronous
dynamics.

Interestingly, our numerical results also show that spatial
correlations do exist at all times and only at the steady state do
they strictly vanish.  It is straightforward to convert the
steady-state equation \eqref{Pw-eq} into a discrete-time recursion
equation for the weight density. Such recursion equation implies fast
exponential relaxation toward the steady state, $dM_k/dn\sim
\exp[-(1-u_k)n]$ for $k\geq 2$. However, our simulations reveal slow
algebraic relaxation instead (figure \ref{fig-m23}) $ dM_k/dn\sim
n^{-3/2}$, for $k\geq 2$.  Spatial correlations, which steadily
diminish with time and eventually disappear altogether, are
responsible for slow relaxation toward the steady state.  The
diffusive relaxation we observe numerically is consistent with the
first-passage behavior \eqref{QT}, and is reminiscent of
time-dependent behavior in reaction-diffusion processes involving
aggregation in one spatial dimension \cite{jls,brt,bdb}.

\bigskip
The authors are grateful to the Kavli Institute for Theoretical
Physics for support, where this research was supported in part by the
National Science Foundation under Grant No. PHY11-25915.

\end{document}